\documentclass{iau} 
\usepackage{graphicx}

\title[Magnetar nebulae can be rotationally powered] 
{Magnetar nebulae can be rotationally powered}

\author[D. F. Torres]   
{Diego F. Torres$^1$
}

\affiliation{$^1$Institute of Space Sciences (IEEC-CSIC), Campus UAB, Carrer de Magrans s/n, 08193 Barcelona, Spain \\ email: {\tt dtorres@ice.csic.es} \\[\affilskip]
$^2$Instituci\'o Catalana de Recerca i Estudis Avan\c{c}ats (ICREA), E-08010 Barcelona, Spain}

\pubyear{2017}
\volume{337}  
\jname{Pulsar Astrophysics - The Next 50 Years}
\editors{P. Weltevrede, B.B.P. Perera, L. Levin Preston \& S. Sanidas, eds.}

\begin{document}

\maketitle

\begin{abstract}
A wind nebula generating extended X-ray emission was recently detected surrounding Swift 1834.9-0846. This is  the first magnetar for which such a (pulsar) wind nebula (PWN) was found. I demonstrate that Swift 1834.9-0846's nebula can be rotationally-powered if it is being compressed by the environment. The physical reason behind this is the dominance of adiabatic heating over all other cooling and escape processes. This effect can happen only for pulsars of relatively low spin-down power and can make for very efficient nebulae. This contribution is based on previous work published in ApJ 
835, article id. 54, 13 pp. (2017).
\end{abstract}

\firstsection 
\section{Introduction}

Swift J1834.9-0846 
was found by the Swift X-ray satellite
on 2011 August 7, via a short X-ray burst. It has a spin period P = 2.48 s and a dipolar magnetic field in the magnetar range (B=1.1$\times 10^{14}$ G, see Younes et al. 2012 for details). Its distance was estimated to be 4 kpc (Esposito et al. 2012)
The spin-down power derived from the measured timing parameters is relatively high for magnetars (2$\times 10^{34}$ erg s$^{-1}$), although not unique.
Observations in quiescence revealed it is surrounded by extended ($\sim$3 pc) X-ray emission (Younes et al. 2012). The nebula luminosity is $L \sim $2$\times 10^{33}$ erg s$^{-1}$, implying a 10\% efficiency (to be compared with the typical efficiencies for PWNe, around 2\%).

Younes et al. (2016) too recently reported on 
extended emission centered at the magnetar. This emission is slightly asymmetrical
and non-variable in a period of ten years (2005-2015). 
Since an interpretation via 
scattering of soft X-ray photons by dust is unfavored due to the constancy of the flux and the hardness of the X-ray spectrum ($\Gamma$ = 1 -- 2), this extended emission was proposed to be the first detection of a magnetar nebula.

We know low-field magnetars (e.g., Rea et al. 2014), and radio emission from magnetars (e.g., Anderson et al. 2012).
We detected magnetar-like bursts from normal pulsars (e.g., Kennea et al. 2016), some have a PWN.
The magnetar's radio emission can be powered by the same physical mechanism responsible for the radio emission in other pulsars (Rea et al. 2012). 
Thus, the existence of a rotationally-powered magnetar nebula would only emphasize the connection between all pulsar classes.

Whereas the existence of nebula surrounding a magnetar would not lead to surprise per se, 
a nebula that powerful and that large, from a pulsar that dim does.
Are 2$\times 10^{34}$ erg s$^{-1}$ of total energy reservoir enough to power a nebula 3 pc in size that emits 2$\times 10^{33}$ erg s$^{-1}$ only in X-rays?
How can a rotationally-powered, ``normal" nebula do it?  

\section{Model}

To answer this we use fully time-dependent PWNe with a detailed expansion model.
Dynamical details of such model can be found in Martin, Torres \& Pedaletti (2016). See Martin, Torres, \& Rea (2012) for 
formulae related to the computation of losses. 
This model computes the evolution in time of the pair distribution within the PWN, subject to 
synchrotron, inverse Compton, and Bremsstrahlung interactions, adiabatic losses, and accounting for escaping particles. 
The model also contains a detailed analysis of reverberation. 
Reverberation occurs when the PWN shell goes into the shocked medium of the remnant and starts the compression. During this phase, the magnetic field and the internal pressure increase.

\section{Size constraints}

\begin{center}
\begin{figure}
\includegraphics[scale=0.53]{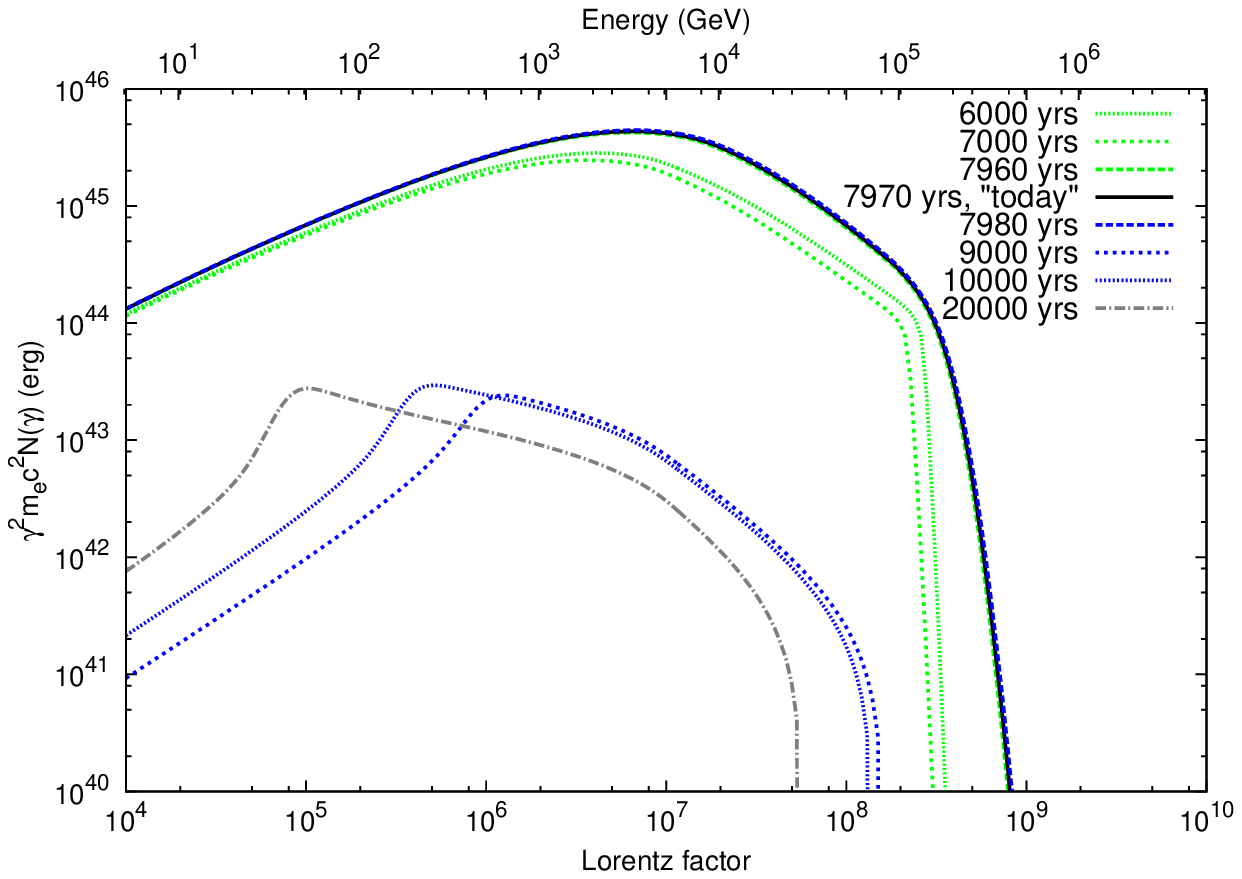} 
\includegraphics[scale=0.53]{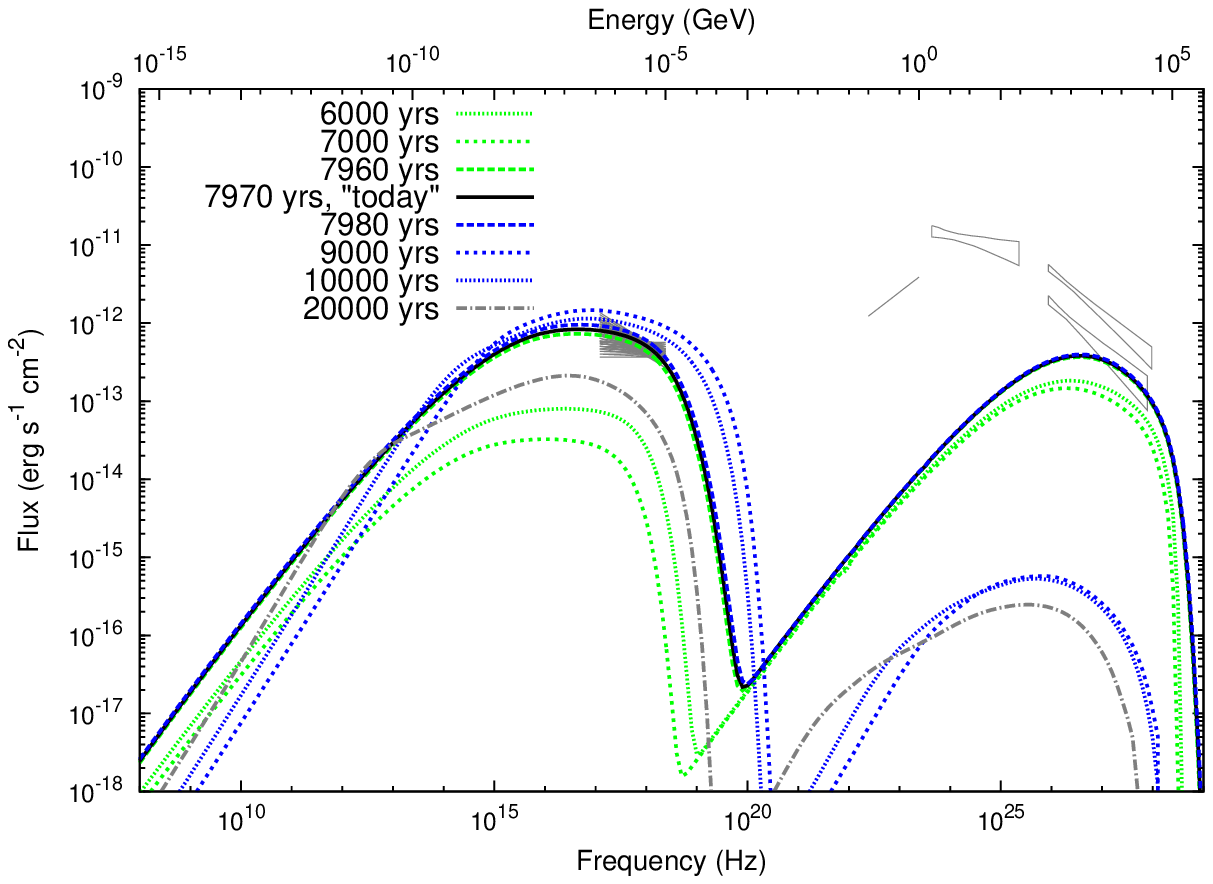} 
\caption{In black in both panels: matching solution today, at 7970 yrs of age for a rotationally-powered magnetar nebula at the start of the reverberation phase. Ten years before or after that age, the nebula would be practically unchanged. The left (right) panel shows the electron population (spectral energy distribution) evolution in time. Both panels shows the dramatic changes of the nebula along the reverberation process.}
\label{7900-TE}
\end{figure}
\end{center}

The measured size of the X-ray nebula  implies constraints on the age (the characteristic age $\tau_c$ of the pulsars is 4.9 kyr). 
If the age is too small (0.6, 1.0 $\tau_c$), the pulsar would be too young to be free-expanding a rotationally-powered nebula up to the size detected.
If the age is too large ($>1.6$ $\tau_c$), the PWN expansion would have been already stopped by the medium and even when re-expanding, its size would be smaller than detected. Also, other problems would appear: low numbers of high energy electrons are left alive after reverberation.
Solutions matching the nebula radius have an age $\sim $1.6 $\tau_c$ , at the end of the free expansion or the beginning of the compression phase, depending on assumptions on the environmental variables. At this age,  the nebula has not compressed too much by the reverberation process. 
Can some of these solutions lead to a good spectral matching?
The answer is yes, as shown in Fig. \ref{7900-TE}, and the model parameters (e.g., magnetization is at 4.5\%) are similar to all other nebulae known (see Torres et al 2017 for details). The only difference, if any, is a slightly larger energy break at injection (at a Lorentz factor of 10$^{7}$, which finds a natural explanation, see below).  The age of the pulsar in this fit is 7900 years, and it can be seen that during the last 10 years, the X-ray flux would have been practically constant. Beyond that time, both in the future and in the past, the time evolution of the nebula is important.

So, where is the trick?
How can the nebula be so efficient? 
At the derived age of the magnetar (7900 years), it is reverberating. 
The PWN size is quickly decreasing.
The $B$-field is quickly increasing.
The PWN pressure is also quickly increasing.
All this is expected in all reverberation processes.
The system will be more luminous first, until most of the high-energy electron population are burned off and re-expansion starts. 
In this solution, the bounce will happen at about 8370 yrs. 
The electron distribution will significantly decrease in the re-expansion of the nebula (see the population at 9000 yrs and beyond). 
However,  at the beginning of the re-expansion, the concurrent high-value of
the magnetic field due to the earlier compression (it is still at about 200 $\mu$G at 9000 yrs, for instance) will make for a sustainable synchrotron power. 
This explains why the level of the X-ray part of the SED is maintained or even increases up to this time, despite the significant reduction in the number of pairs. 
Such magnetic effect does not apply
to the inverse Compton yield, which decreases all along.
When pairs are burned off at the start of the Sedov phase, while the high-energy particles are slowly building up only by injection, the synchrotron luminosity will finally decrease (see the SED at 20000 yrs). 
The radius that this future nebula will have is smaller than 1 pc, having been affected by the strong compression process, which goes almost unimpeded due to the relatively low pulsar power.

It is interesting to note too what has happened before reaching the current age. 
Fig. \ref{7900-TE} compares the current SED and electron spectrum with their values 
at 6000 and 7000 yrs. 
At both of these ages, the pulsar was still free expanding. 
The larger the age (at 7000 vs. 6000 yrs), the PWN was larger (5.0 vs 4.3 pc), the magnetic field was smaller (1.65 vs 2.25 $\mu$G), and 
more particles were affected by losses, for which the timescales started to be of comparable magnitude to the age (see next). 
Thus, the electron population is larger at 6000 yrs than at 7000 yrs, and the SED is correspondingly more luminous.
What happens next, between 7000 yrs and today, depends strongly on the reverberation process, and constitutes
the reason why the nebula can actually be observed.
In pulsars of low spin-down power reverberation induces a large increase in the number of high-energy particles, by heating the pool of particles injected by the pulsar at earlier times. 
That heating happens due to the adiabatic transfer of energy from the environment to the nebula, and 
reflects in the spectral energy distribution since suddenly (in a period of few hundred years) there are plenty more particles able to emit in X-rays.

\begin{center}
\begin{figure}
\centering
\includegraphics[scale=0.5]{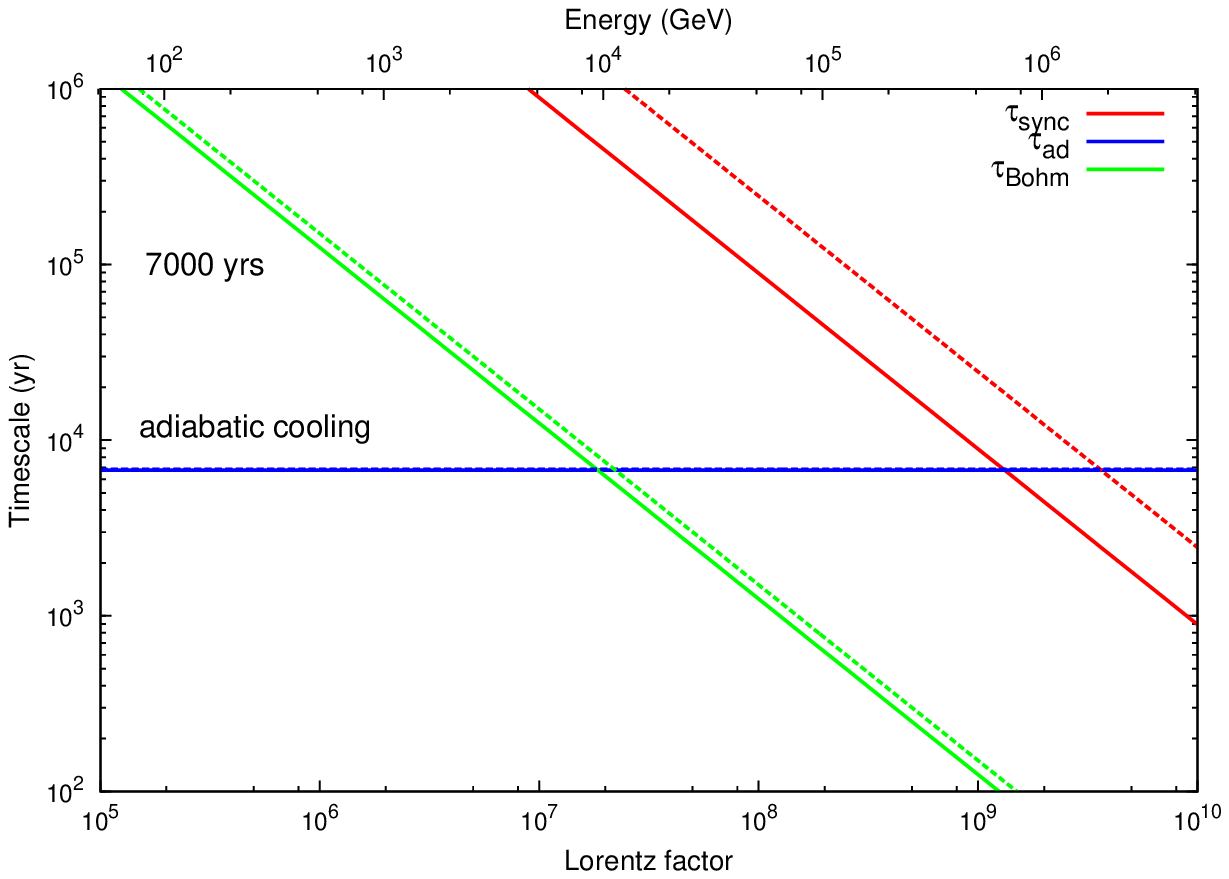}
\includegraphics[scale=0.5]{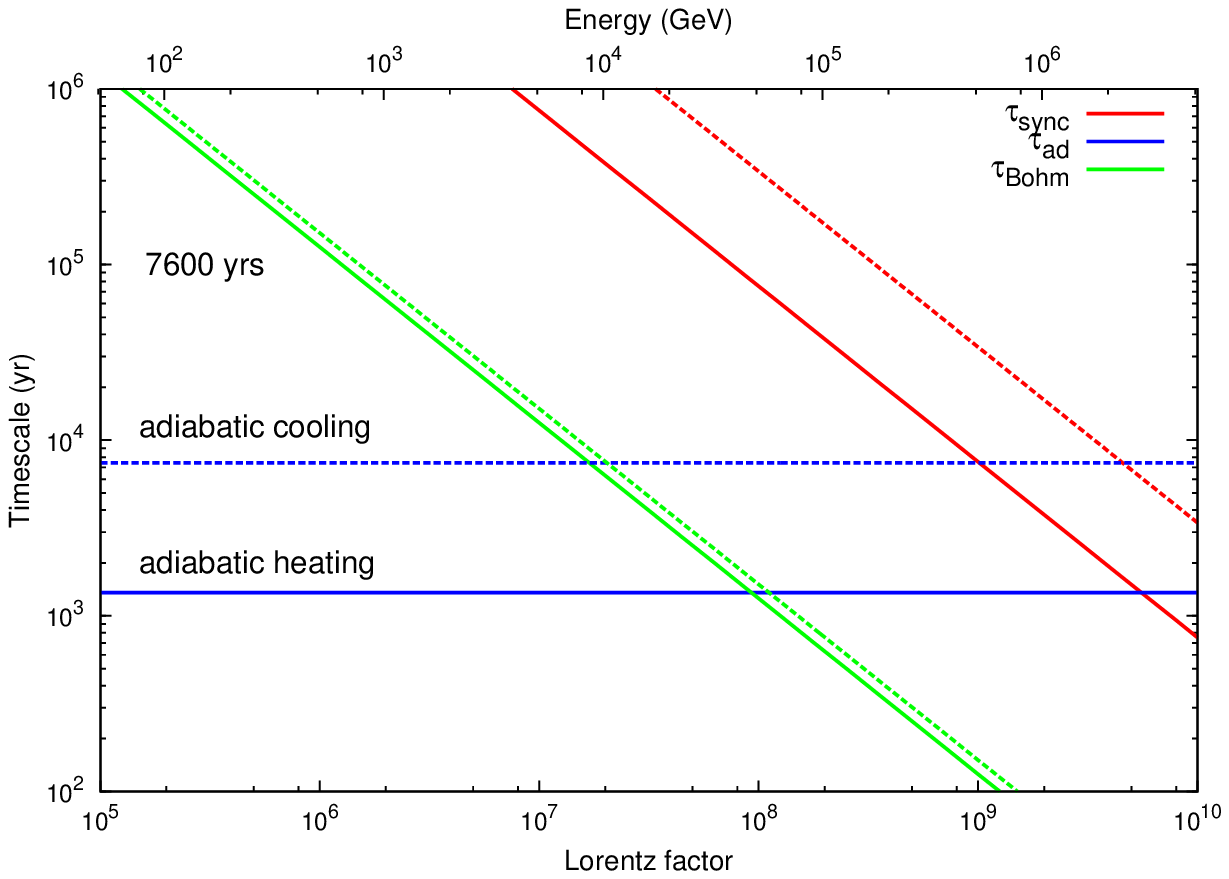}\\
\includegraphics[scale=0.5]{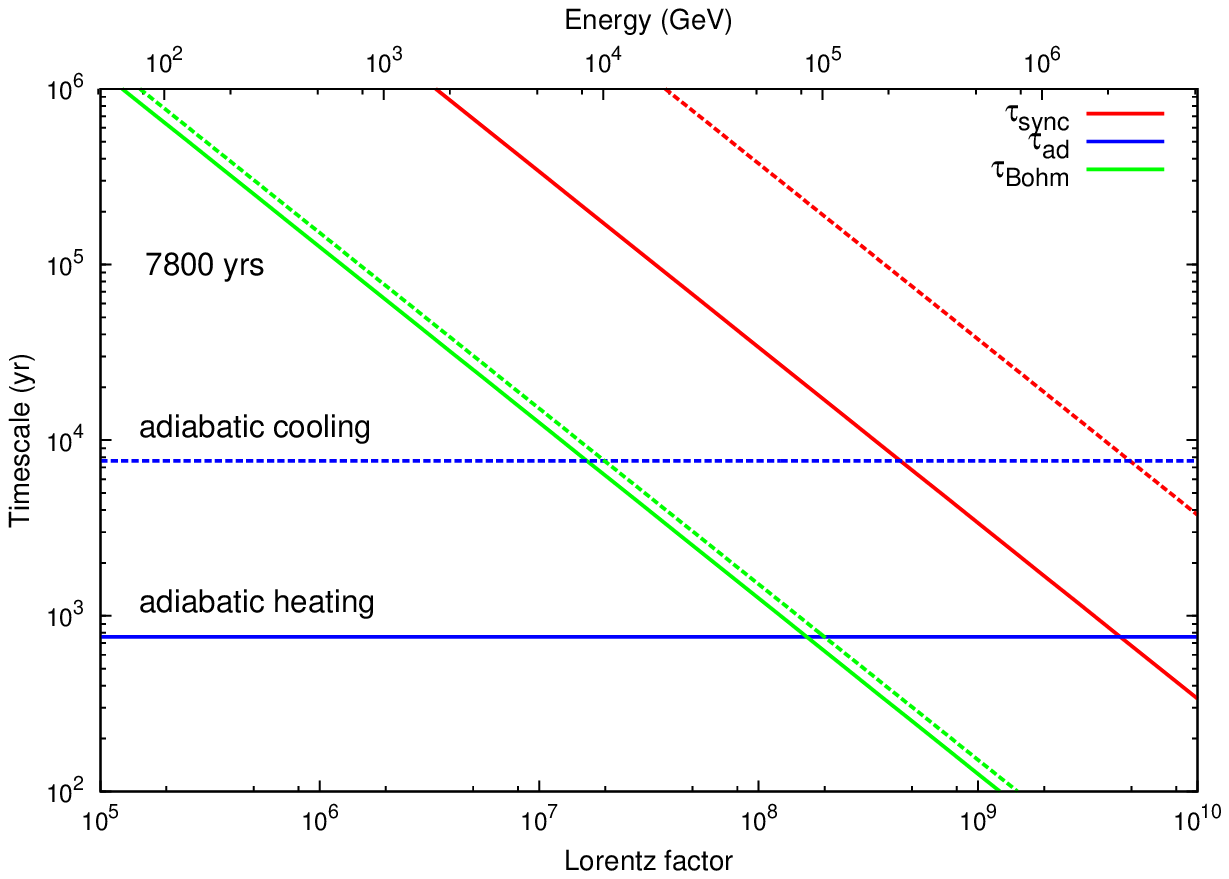}
\includegraphics[scale=0.5]{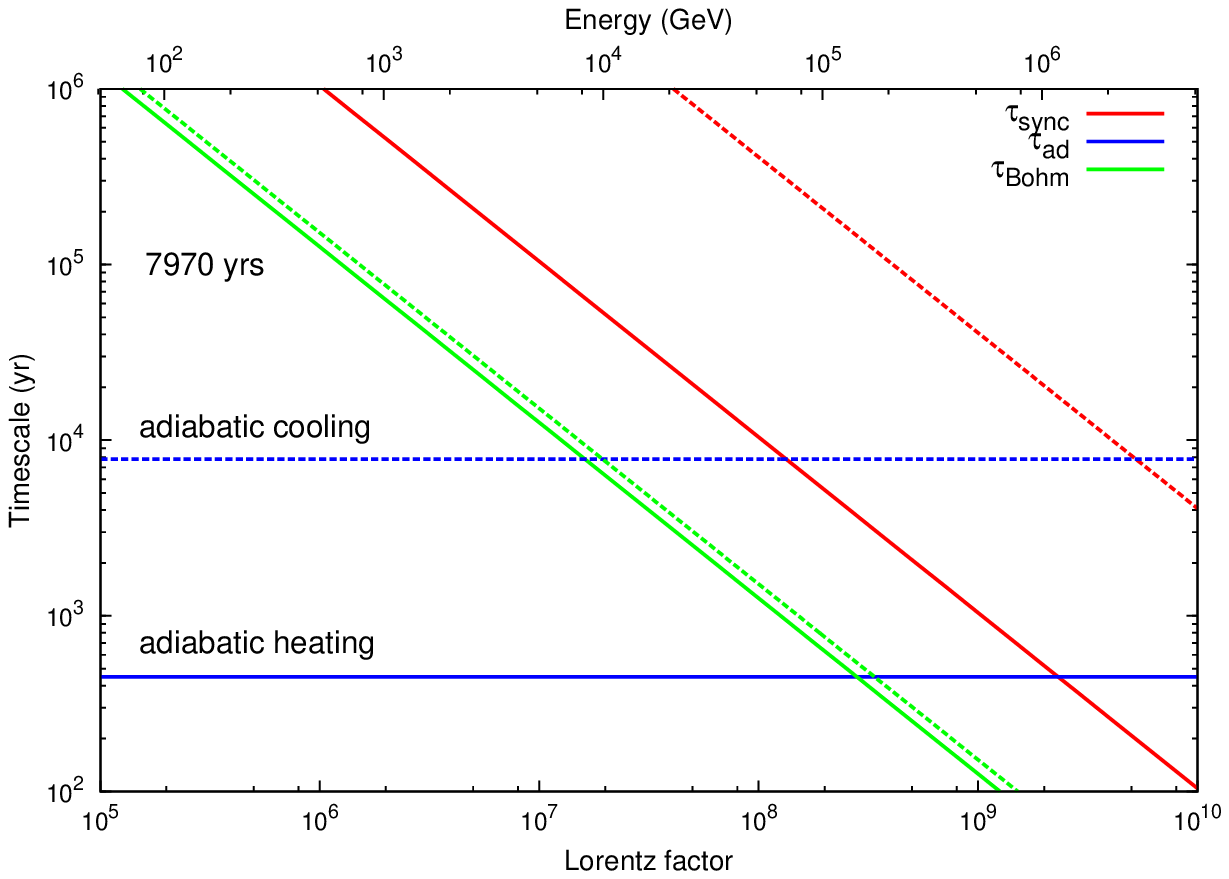}
\caption{Comparison of timescales for relevant particle losses (and energy gain in the case of the adiabatic timescale along reverberation) for models of age and parameters equal  to the matching one shown in Fig. 1. In order to compare the importance of reverberation, shown are models with dynamical evolution without (dashed lines) and with (solid lines) reverberation being considered. Subdominant bremsstrahlung and inverse Compton timescales are not shown for clarity but also considered in the computation.
}
\label{comp1}
\vspace{0.2cm}
\end{figure}
\end{center}

Fig. \ref{comp1} shows the losses/heating and escape timescales at different times. The smaller is a given timescale, the more dominant the corresponding process is with respect to the pair population.
We see how for the current age and since a few hundred years earlier, the heating timescale dominates and allows particles to become X-ray emitters when reverberation is ongoing.
Such a dominance of the heating over the losses can only occur in pulsars of low spin down.
Like in a magnetar.
In them, because the spin-down power is relatively low, so is the magnetic pressure and the synchrotron losses are far from dominating the energy balance. 
In more energetic pulsars, adiabatic heating is not so intense,
the reverse shock velocity is smaller, and synchrotron losses dominate.

\section{Conclusions}

Magnetar nebulae can be rotationally-powered. An additional source of energy is not needed in order to understand the observations of the first nebula detected around Swift 1834.9-0846.
 The requirement for this to happen 
is that the nebula is currently 
being compressed by the environment. 
We have found that this is possible for an age of around 8000 years, about 1.6 times the estimated characteristic age. 
Our findings imply that  the spin-down (and related) parameters are not markers of the PWN detectability, and that even super-efficient nebula (where the X-ray luminosity exceeds the spin-down) are possible without violating any energetic constraint.

\acknowledgments

I acknowledge support from the grants AYA2015-71042-P, and SGR 2014-1073.

\end{document}